\documentclass[pre,twocolumn,aps,eqsecnum]{revtex4}
\usepackage{epsfig}

\begin{document}

\title{Anomalous diffusion in heterogeneous glass-forming liquids: Temperature-dependent behavior}

\author{ J. S. Langer}
\affiliation{Dept. of Physics, University of California, Santa
Barbara, CA  93106-9530}

\date{\today}

\begin{abstract}
In a preceding paper \cite{JSL-SM-08}, Mukhopadhyay and I studied the diffusive motion of a tagged molecule in an heterogeneous glass-forming liquid at temperatures just above a glass transition. Among other features of this system, we postulated a  relation between heterogeneity and stretched-exponential decay of correlations, and we also confirmed that systems of this kind generally exhibit non-Gaussian diffusion on intermediate length and time scales.  Here I extend this analysis to higher temperatures approaching the point where the heterogeneities disappear and thermal activation barriers become small. I start by modifying the continuous-time random-walk theory proposed in \cite{JSL-SM-08}, and supplement this analysis with an extension of the excitation-chain theory of glass dynamics.  I also use a key result from the shear-transformation-zone theory of viscous deformation of amorphous materials.  Elements of each of these theories are then used to interpret experimental data for ortho-terphenyl, specifially, the diffusion and viscosity coefficients and neutron scattering measurements of the self intermediate scattering function. Reconciling the theory with these data sets provides insights about the crossover between super-Arrhenius and Arrhenius dynamics, length scales of spatial heterogeneities, violation of the Stokes-Einstein relation in glass-forming liquids, and the origin of stretched exponential decay of correlations.  

\end{abstract}

\maketitle

\section{Introduction}
\label{intro}

Many of the most important questions in the physics of glassy materials have to do with spatial heterogeneities.  What is the scale of these heterogeneities?  In what sense are they dynamic as opposed to structural phenomena?  What is their relation to stretched exponential decay of correlations?  To anomalous, non-Gaussian diffusion?  To violations of the Stokes-Einstein relation between diffusivity and viscosity?  I address each of these issues in this paper by studying a relatively simple model in which all of them arise, and which allows me to make contact with experimental data.  

In a preceding paper \cite{JSL-SM-08}, Mukhopadhyay and I studied a model of the diffusive motion of a tagged molecule in a heterogeneous glass-forming liquid.  This model is based in part on the excitation-chain theory of the glass transition.\cite{Langer-PRE-06,Langer-PRL-06} It consists of two different kinds of molecular environments that we called ``glassy'' and ``mobile.''  To a first approximation, the tagged molecule is either frozen at a fixed position in a glassy domain or is moving diffusively in a mobile region.  The glassy domains are compact regions of relatively low energy and entropy, surrounded by more highly disordered regions in which molecular rearrangements can occur.  Because of the mobility of the molecules within and at the boundaries of the mobile regions, these boundaries diffuse slowly throughout the system on an ``$\alpha$'' relaxation time scale, and the system as a whole is ergodic.  A tagged molecule in a glassy domain is stationary until it is encountered by a moving boundary, at which time its motion is enabled and it diffuses until it finds itself back in a glassy domain.  In this sense, the model has elements of kinetically constrained systems. \cite{FREDRICKSON85,FREDRICKSON86,KOB93,TONINELLI04,GARRAHAN-CHANDLER02,GARRAHAN-CHANDLER03,GARRAHAN-CHANDLER05}

The analysis in \cite{JSL-SM-08} pertained only to behavior at temperatures $T$ low enough that the  heterogeneities are much larger than the molecular spacing.  With that assumption, we found a clear decoupling between the slow $\alpha$ relaxation and the faster mobile motions that we identified as ``$\beta$'' cage-breaking rearrangements.  Fluctuations in the sizes of the glassy domains produced stretched-exponential decay of the self intermediate scattering function at long times, and the distribution over spatial displacements exhibited an anomalously broad exponential (non-Gaussian) tail during the transition from $\beta$- to $\alpha$-like motion.  

My purpose here is to reformulate \cite{JSL-SM-08} in a way that explores the decrease in the size of the heterogeneities as a function of increasing temperature above the glass transition.  My ultimate goal is to understand the crossover between the thermally activated phenomena that occur near the glass temperature and the non-activated, fluidlike transport that is described at high temperatures by mode-coupling theory.  

I first reformulate the preceding analysis in Sec.\ref{CTRW}, pointing out how the theory must be modified as the system changes from being predominantly glasslike at low temperatures to predominantly mobile near the transition from super-Arrhenius to Arrhenius dynamics at a crossover temperature $T_A$. Then, in Sec.\ref{XC}, I review the excitation-chain theory \cite{Langer-PRE-06,Langer-PRL-06} of these temperature-dependent quantities and show how, in a simple approximation, the basic ideas of that theory may be extended to describe the behavior near the crossover.  

The resulting temperature-dependent scattering functions and diffusion and viscosity coefficients are compared with experimental data for ortho-terphenyl (OTP) in Sec.\ref{OTP}.  Among other results, this analysis makes it appear that the spatial heterogeneities in OTP are relatively small, and that the observed violation of the Stokes-Einstein relation near the glass temperature occurs because the mechanisms responsible for diffusion and viscosity become different from one another as the glass-forming liquid  becomes more like a solid.  Section \ref{secSED} contains a discussion of the way in which stretched exponential decays of correlations emerge in a theory of this kind.  Finally, in Sec.\ref{conclusions}, I summarize my present opinions about the issues raised in the first paragraph of this Introduction. 

\section{Continuous-time random walks}
\label{CTRW}

The continuous-time random-walk (CTRW) analysis \cite{MONTROLL-SHLESINGER-84,BOUCHAUD-GEORGES-90} used in \cite{JSL-SM-08} required that  molecular motions consist of just two alternating modes -- glassy and mobile -- each with its own characteristic time scale.  The present analysis is based on a similar two-stage assumption, modified in a way that more naturally crosses over to high-temperature behavior in which the tagged molecule undergoes only normal diffusion. The two-stage assumption is a serious one because it precludes exploring molecular motions in the continuous range of short time scales between  molecular vibration periods and $\beta$ relaxation, i.e. in the so-called ``ballistic'' regime.  This analysis cannot resolve motions on those very short time scales, or on length scales appreciably smaller than the intermolecular spacing.  Nevertheless, the CTRW technique is uniquely capable of describing the larger scale motions which do exhibit a clear decoupling between fast and slow modes. The power of this kind of analysis is illustrated in \cite{CHAUDHURIetal07}, which has served as a starting point for much of the present work.

With this short-time limitation in mind, denote the two temperature-dependent time scales by $\tau_{\alpha}(T)$ for the long, $\alpha$ relaxation time, and $\tau_M(T)$ for the ordinarily shorter, mobile relaxation time.  As in \cite{JSL-SM-08}, let the characteristic size of the heterogeneities be $\ell\, R^*(T)$, where $\ell$ is a length of the order of the average molecular spacing and $R^*$ is a dimensionless function of $T$.  Then the characteristic fluctuation time for the heterogenities is
\begin{equation}
\label{tau*def}
\tau^R_{\alpha}(T) =  {R^*}^2(T)\,\tau_{\alpha}(T).
\end{equation}

A central ingredient in this analysis is the probability ${\cal P}_G(T)$ that the tagged moleecule is in a glassy domain.  This probability is a geometrical quantity, equal to the fraction of the system occupied by glassy domains, which is assumed to be approximately a constant in time although the positions of the domains themselves are fluctuating.  

The waiting-time probability distribution for a molecule initially in a glassy domain was found in \cite{JSL-SM-08} to be
\begin{equation}
\label{psiG}
\psi_G(t^*)= 2\,e^{-2\,\sqrt{t^*}},
\end{equation}
where $t^*=t/\tau_{\alpha}^R$ and $t$ is the time in dimensional units. Note that $\psi_G(t^*)$ is normalized over $t^*$ rather than $t$.  This distribution, which has the Kohlrausch form $\exp\,(- {\rm const.}\,\times {t^*}^b)$ with a stretched-exponential index $b=1/2$, was obtained by assuming a Gaussian distribution of domain sizes $R$ with a scale size $R^*(T)$.  The index $b$ is probably a temperature dependent quantity, rising to unity as $T$ increases to the crossover temperature $T_A$; but here, as in \cite{JSL-SM-08}, I make no attempt to estimate this $T$ dependence and simply keep $b = 1/2$, independent of $T$.  As will be shown in Sec.\ref{secSED}, this ``intrinsic'' value of $b$ is generally not what is observed experimentally.   

The main departures from \cite{JSL-SM-08} that are needed here concern molecular motions in the mobile regions.  The revised model must make a smooth transition between the low and high temperature regimes.  At low $T$, the analysis in \cite{Langer-PRE-06,Langer-PRL-06} suggests that the system consists primarily of glassy domains surrounded by thin, tenuous mobile regions.  When a molecule enters a mobile region under these conditions, I assume that it moves a distance proportional to $R^*$ and, with high probability, is back in a glassy domain at the end of this motion.  By letting this jump-length scale with $R^*$, the size of the heterogeneities, I account roughly for fact that the mobile regions may extend over large distances, even though they occupy only a small fraction of the total volume at low temperatures.

In contrast, at higher temperatures near $T_A$, the system consists primarily of mobile regions in which small glassy domains are embedded.  In this case, I assume that a molecule makes multiple diffusive jumps in the mobile region, and only with low probability finds itself in a glassy domain after any single jump.  A constraint on the model is that, as $T$ approaches and exceeds $T_A$, molecular motion should cross over from anomalous to normal diffusion with a diffusion constant $D_M \propto\ell^2/\tau_M$.  

Because no stretched-exponential behavior is expected to be associated with diffusion in mobile regions, let the waiting time between mobile jumps have a simple exponential distribution:
\begin{equation}
\label{psiM}
\psi_M(t^*) = \Delta\,e^{-\Delta\,t^*},
\end{equation} 
where $\Delta$ is the dimensionless ratio between the glassy and mobile residence times.  There are various, equally plausible choices for $\Delta$, the simplest of which is
\begin{equation}
\label{Deltadef}
\Delta = \tau_{\alpha}/\tau_M.
\end{equation}
With this choice, and with the definition of $t^*$, the characteristic mobile waiting time is of the order of $\tau_M\,{R^*}^2 \propto \ell^2\,{R^*}^2/D_M$.  If the time between jumps scales like the square of the domain size, $(\ell\,R^*)^2$, then the characteristic jump length must scale like $\ell\,R^*$; therefore write the jump-size probability in the form 
\begin{equation}
\label{frjump}
f(r^*) = {1\over (2\,\pi\,a^2)^{d/2}}\,\exp\,\left(-{{r^*}^2\over 2\,a^2}\right),
\end{equation} 
where $r^*= r/(\ell\,R^*)$ is the jump length $r$ in units of $\ell\,R^*$, $a$ is a dimensionless parameter, and $d$ is the dimensionality. (Ordinarily $d=3$ in the examples to be shown here. For some purposes such as in Sec. \ref{XC}, however, it is useful to keep general $d$.)  

This picture so far is consistent with the idea that, when $R^*$ is large at low $T$, the jump size and the waiting time are both large.  At high $T$ however, where $R^*$ is small, the picture makes sense only if multiple jumps can occur in mobile regions.  Thus, assume that a mobile jump is followed by another mobile jump with probability ${\cal P}_M = 1 - {\cal P}_G$, and that a molecule ends in a glassy domain after such a jump with probability ${\cal P}_G$.  This assumption means that, at low $T$ where ${\cal P}_G$ is nearly unity and ${\cal P}_M$ is small, the molecule makes mostly infrequent, long jumps; whereas, at high $T$ where ${\cal P}_M$ is close to unity and ${\cal P}_G$ is small, the molecule makes repeated, fast, short jumps.  
 
The mathematics of the CTRW analysis is much the same here as in \cite{JSL-SM-08}.  As usual, it is easiest to compute with the Fourier-Laplace transforms of the constituent probability distributions. Define two different distribution functions, $n_G(r^*,t^*)$ and $n_M(r^*,t^*)$, for molecules starting, respectively, in glassy domains or mobile regions, and moving distances $r^*$ in times $t^*$. Their Fourier-Laplace transforms are
\begin{equation}
\tilde n_j(k,u)=\int e^{-i{\bf k^*}\cdot{\bf r^*}}d{\bf r^*}\int_0^{\infty} e^{-ut^*} n_j(r^*,t^*)\,dt^*,
\end{equation}
where $j = G,M$, ${\bf k^*} = {\bf k}\,\ell\,R^*$, and ${\bf k}$ is the dimensional Fourier wave vector. 

The Laplace transform of the glassy waiting time distribution $\psi_G(t^*)$ is
\begin{equation}
\label{psiGu}
\tilde \psi_G(u)=\int_0^{\infty} e^{-ut^*}\psi_G(t^*)\,dt^*.
\end{equation}
As discussed in \cite{JSL-SM-08}, this function has an essential singularity at $u=0$ and a branch cut conveniently chosen to lie along the negative real $u$ axis.  The probability that the molecule has not yet left its glassy domain at time $t^*$ is
\begin{equation}
\label{phiG}
\phi_G(t^*)=\int_{t^*}^{\infty} \psi_G({t^*}')\,d{t^*}'= (1+2\,\sqrt{t^*})\,e^{-2\,\sqrt{t^*}};
\end{equation}
and the Laplace transform of this function is 
\begin{equation}
\label{phiGu}
\tilde\phi_G(u)={1-\tilde\psi_G(u)\over u}.
\end{equation}

The analogous functions for mobile motion are
\begin{equation}
\tilde \psi_M(u) = {\Delta\over u+\Delta};~~~~\tilde\phi_M(u) = {1\over u+\Delta}.
\end{equation}
The Fourier transform of the jump-length probability distribution in Eq.(\ref{frjump}) is
\begin{equation}
\label{fkjump}
\hat f(k^*)= \exp\,\left(- {a^2\,{k^*}^2\over 2}\right),
\end{equation}
where $k^*=|{\bf k}^*|$.

The next step in the analysis is to consider a sequence of jumps in which a molecule starts in a mobile region and does not enter a glassy domain until its last jump.  The corresponding propagator is 
\begin{equation}
{\cal F}(k^*,u)={{\cal P}_G\,\hat f(k^*)\,\tilde\psi_M(u)\over 1 - {\cal P}_M\,\hat f(k^*)\,\tilde\psi_M(u)}.
\end{equation}
A related function describes the sequence in which the molecule remains in the mobile region after its last jump:
\begin{equation}
{\cal G}(k^*,u)={\tilde\phi_M(u)\over 1 - {\cal P}_M\,\hat f(k^*)\,\tilde\psi_M(u)}.
\end{equation}
With these ingredients, the functions $\tilde n_G(k^*,u)$ and $\tilde n_M(k^*,u)$ become
\begin{eqnarray}
\label{nGku}
\nonumber
\tilde n_G(k^*,u)&=& {\tilde \phi_G(u)+{\cal G}(k^*,u)\,\tilde \psi_G(u)\over 1-{\cal F}(k^*,u)\,\tilde \psi_G(u)}
\\ \cr &=&{1\over u}\,{N_G(k^*,u)\over W(k^*,u)};
\end{eqnarray}
and
\begin{eqnarray}
\label{nMku}
\nonumber
\tilde n_M(k^*,u)&=& {\tilde \phi_G(u)\,{\cal F}(k^*,u)+ {\cal G}(k^*,u)\over 1-{\cal F}(k^*,u)\,\tilde \psi_G(u)}
\\ \cr &=&{1\over u}\,{N_M(k^*,u)\over W(k^*,u)};
\end{eqnarray}
where
\begin{equation}
N_G(k^*,u) = \left[1-{\cal P}_M\,\hat f(k^*)\right]\left[1-\tilde \psi_G(u)\right] + u/\Delta;
\end{equation}
\begin{equation}
N_M(k^*,u)= \left[1-\tilde\psi_G(u)\right]\,{\cal P}_G\,\hat f(k^*)+u/\Delta;
\end{equation}
and
\begin{equation}
\label{Wdef}
W(k^*,u)= 1 - {\cal P}_M\,\hat f(k^*) +u/\Delta - {\cal P}_G\,\hat f(k^*)\,\tilde\psi_G(u). 
\end{equation}
In Eqs.(\ref{nGku}) and (\ref{nMku}), the first term in the numerator corresponds to the case in which the molecule ends in a glassy domain and the second term to the case where it ends in a mobile region.  

The Laplace transformed self intermediate scattering function, $\tilde F_s(k^*,u)$, is the weighted average of $\tilde n_G$ and $\tilde n_M$:
\begin{equation}
\label{ISFudef}
\tilde F_s(k^*,u)= {\cal P}_G(T)\,\tilde n_G(k^*,u)+ {\cal P}_M(T)\,\tilde n_M(k^*,u).
\end{equation}
The scattering function that is measured experimentally is the inverse Laplace transform:
\begin{equation}
\label{ISFtdef}
\hat F_s(k^*,t^*)=\int_{- i\infty}^{+ i\infty} {du\over 2\,\pi\,i}\,e^{u\,t^*}\,\tilde F_s(k^*,u).
\end{equation}
As in \cite{JSL-SM-08}, this integration is performed by closing the contour around the branch cut on the negative real $u$ axis.  The crucial ingredient is, for $u\to -w\pm i\epsilon$, $\epsilon \to +0$,  
\begin{eqnarray}
\label{A+iB}
\nonumber
&&\tilde \psi_G(-w\pm i\epsilon)=-4\int_0^{\infty}\xi\,d\xi\,e^{-w \xi^2\pm 2 i \xi}\cr &&= 
-{2\over w}+{2\sqrt{\pi}\over w^{3/2}}\,e^{-1/w}\left({\rm Erfi}\Bigl({1\over\sqrt{w}}\Bigr)
\mp \,i\right)\\ \cr&&\equiv A(w) \mp i\,B(w),
\end{eqnarray}
where Erfi is the imaginary error function.  The imaginary part of $\tilde \psi_G(-w\pm i\epsilon)$ is its discontinuity across the branch cut on the negative $u$ axis. With this formula, it is straightforward to compute the discontinuity across the cut for the functions $\tilde n_j(k^*,u),~~ j=G,\,M$ in Eqs.(\ref{nGku}) and (\ref{nMku}), and in this way to compute the scattering functions by closing the contour around this cut. That is:
\begin{eqnarray}
\label{hatnj}
\nonumber
\hat n_j(k^*,t^*)&=& \lim_{\epsilon \to +0}\int_0^{\infty}dw\,{e^{-w\,t^*}\over \pi\,w}\\ \cr &\times&{\rm Im} \left[{N_j(k^*,-w+i\epsilon)\over W(k^*,-w+i\epsilon)}\right],~~j=G,\,M.~~~~~~~
\end{eqnarray}
The integrands are well behaved as $w \to +0$, therefore it is possible to let the lower limit of integration be $w=0$.

\section{Excitation-chain theory}
\label{XC}

Evaluation of the formulas in Sec.\ref{CTRW} requires explicit expressions for the temperature-dependent quantities $\tau_{\alpha}(T)$, $\tau_M(T)$, $R^*(T)$, and ${\cal P}_G(T)$.  These expressions are needed not just as asymptotic forms near the glass transition but, more importantly, as physically motivated approximations valid near the crossover from anomalous to normal diffusion near $T_A$.  That crossover coincides with a transition from super-Arrhenius to Arrhenius dynamics, and from heterogeneous to homogeneous spatial properties. 

I propose to use the excitation-chain (XC) theory \cite{Langer-PRE-06,Langer-PRL-06} to derive these formulas.  So far as I know, it is at present the only theory of glass-forming liquids that relates their dynamic and thermodynamic properties directly to the behavior of real molecules with short-range interactions, and that does so in a manner consistent with the basic principles of statistical mechanics.  This theory remains speculative and incomplete.  It does not even tell us whether the glass transition is thermodynamically well defined or whether, as I strongly suspect, relaxation times grow increasingly rapidly but smoothly all the way down to zero temperature.  If the latter is true, then the XC theory describes just the onset of that process and, in doing so, gives us just a clue about what energy scales and collective motions might be involved.  In either case, it provides a framework for understanding what might be happening near $T_A$.  

In order to go beyond the immediate vicinity of the glass temperature, I need to extend the published XC results, and therefore need to provide a brief summary of the basic ingredients of this theory.  The fundamental question to be addressed is why molecular rearrangements that create or annihilate density fluctuations are not simply thermally activated Arrhenius processes.  My  answer is that simple, one-step, activated processes do occur, but that they do not always lead to stable, configurational rearrangements.  Think of the creation of a density fluctuation very roughly as the formation of the glassy analog of a ``vacancy-interstitial pair'' in which a molecule moves out of an energetically favored site to a neighboring, less favored position, perhaps leaving a lower density at the original site and producing a higher density at the new site.  Denote the energetic cost of this move by $k_B\,T_Z$.  Then think about what happens during the next thermal fluctuation.  The energetically favored next move is recombination; the interstitial falls back into the vacancy. At high enough temperatures, however, this energetic advantage is outweighed by the entropic advantage of there being many different next-neighbor sites available.  The vacancy and the interstitial in effect dissociate from each other in a single activated step, and the resulting density fluctuation is an Arrhenius process with a temperature-independent energy barrier.  

At lower temperatures, by definition below $T_A$, dissociation must become a multi-step process.  The simplest such mechanism that I can imagine is a chain-like sequence in which one molecule pushes or pulls its neighbor, that neighbor displaces the next neighbor, etc., finally leaving a vacancy at one end of the chain of displacements and an interstitial at the other.  Chainlike (or stringlike) excitations appear to be ubiquitous in the neighborhood of jamming transitions in amorphous materials.  For example, see the work of Glotzer {\it et al.} \cite{GLOTZER99,GLOTZER00a,GLOTZER00b}.  The energy cost of such an excitation must, to a first approximation, be proportional to the length of the chain; but this cost may be compensated by an entropic advantage; the longer the chain of displacements, the more routes are available for it to take. 

To interpret this picture mathematically, compute the rate of thermal activation of an excitation chain consisting of $N$ molecular steps and extending a distance $R$ in units $\ell$, where $R$ is dimensionless and $\ell$ is the same molecular length scale that was introduced in Eq.(\ref{tau*def}).  This problem is closely related to -- but not exactly the same as -- the problem of computing the probability of $N$-step random walks with extensions $R$.  Write this rate in the form of an attempt frequency, say $\tau_0^{-1}$, times an activation factor $\exp \bigl(-\Delta G(N,R)/k_B\,T\bigr)$, where:
\begin{equation}
\label{DeltaG1}
\Delta G(N,R)= k_B T_Z + N e_0  - k_B T\,\ln W + E_{int}.
\end{equation}
The first term on the right-hand side, $k_B T_Z$, is the bare activation energy discussed above.  The remaining terms describe the statistics of the chain.  However, I omit the diffusion term, proportional to $R^2/2N$, from the usual random-walk analysis because it turns out to be negligible for present purposes.  

In the second term in Eq.(\ref{DeltaG1}), $e_0$, is the activation energy per link in the chain, roughly the energy required to move a pair of molecules far enough away from each other to allow a third molecule to pass between them. 

The third term, $W(N,R)$, is a sum over chain configurations weighted by the disorder of the glassy environment in which these excitations occur.  It has the form:
\begin{equation}
\label{W1}
 \ln\,W(N,R) \approx  \nu\,N - {\pi\,\gamma(T)\over 2}\,R,  
\end{equation}
where $\exp\,(\nu)$ is the number of choices that the successive links in the chain can make at each step.   The function $\gamma(T)$ is the mean-square amplitude of the fluctuations in $e_0/k_B T$;
\begin{equation}
\gamma(T) = \gamma_0\,\left({T_0 \over T} \right)^2.
\end{equation}
$T_0 = e_0/\nu\,k_B$ is the characteristic temperature determined by the energy $e_0$.  With the approximations used here, $T_0$ turns out to be equal to both the Kauzmann and the Vogel-Fulcher temperatures in the XC theory. $\gamma_0/\ell$ is the inverse localization length associated with diffusion in disordered systems.  Because glassy disorder is on molecular scales, $\gamma_0$ seems likely to be about unity; but a first-principles derivation remains one of the most uncertain and speculative parts of this theory. The derivation given in \cite{Langer-PRE-06} almost certainly breaks down at large $N$ and low temperatures. Thus, it seems likely that this calculation makes sense only at temperatures not too close to $T_0$, and that even the existence of a nonzero $T_0$ remains problematic.  

The last term in Eq.(\ref{DeltaG1}), $E_{int}$, is especially important for the present analysis.  This is the energy that makes it unfavorable for the links of the chain to lie near each other.  In \cite{Langer-PRE-06,Langer-PRL-06}, I used Flory's approximation \cite{FLORY53} for the self-exclusion energy of a polymer chain in a $d$-dimensional environment:
\begin{equation}
\label{EintF}
E_{int}^{Flory}(N,R)\approx k_B\,T_{int}\,{N^2\over R^d},
\end{equation}
where the temperature $T_{int}$ sets the scale of this energy up to a dimensionless factor of the order of unity. This approximation is valid, however, only in the limit of large $N$ and $R$; whereas we need a theory that makes sense in both the latter limit and for vanishing values of those quantities near $T=T_A$. For the moment, therefore, rewrite this energy in the form
\begin{equation}
\label{Eint}
E_{int}(N,R)\approx k_B\,T_{int}\,{J(N)\over R^d},
\end{equation}
where the as-yet undetermined function $J(N)$ is equal to $N^2$ in the limit of large $N$ but becomes a much weaker repulsion in the limit $N\to 0$.  

To compute the activation rate from Eq.(\ref{DeltaG1}), first find the value of $R=R^*$ at which $\Delta G(N,R)$ is a minimum.  The resulting function $\Delta G^*(N) = \Delta G(N,R^*)$ is conveniently written as
\begin{equation}
{\Delta G^*(N)\over k_B} = T_Z + T\,\tilde\alpha(T,N),
\end{equation}
where
\begin{eqnarray}
\label{alphatilde}
\nonumber
&&\tilde\alpha(T,N)= -\nu\,N\,\left(1-{T_0\over T}\right)\cr \\&&+ (d+1)\,\left[{\pi\,\gamma(T)\over 2\,d}\right]^{d\over d+1}\,\left[{T_{int}\over T}\,J(N)\right]^{1\over d+1},
\end{eqnarray}
and
\begin{equation}
R^* = \left[{2\,d\,T_{int}\over \pi\,\gamma(T)\,T}\,J(N)\right]^{1\over d+1}.
\end{equation}

If $T>T_0$, and if the second term on the right-hand side of Eq.(\ref{alphatilde}) is sublinear in $N$, then $\tilde\alpha(T,N)$ has a maximum at $N=N^*(T)$ which, in analogy to nucleation theory, is the length of the critically large chain that nucleates a long-lasting  density fluctuation.  When a chain fluctuates to a size larger than $N^*$, it is highly likely to continue growing without bound, and thus to dissociate the vacancy-interstitial pair.  Thus the activation energy for this process is $\Delta G(N^*,R^*)$, and the $\alpha$ relaxation time $\tau_{\alpha}$ is given by
\begin{equation}
\label{alphadef}
\ln \left({\tau_{\alpha}\over \tau_0}\right) = {T_Z\over T} + \alpha(T),
\end{equation}
where $\alpha(T) = \tilde\alpha[T,N^*(T)]$.   

For temperatures low enough that $N^* \gg 1$, the Flory limit in Eq.(\ref{EintF}) is accurate; that is, $J(N)\approx N^2$, so that 
\begin{equation}
\label{R*}
R^*(T)\approx \left[{\pi\,\gamma(T)\,T\,T_{int}\over 2\,d}\right]^{1\over d-1}\,\left[{2\over \nu\,(T-T_0)}\right]^{2\over d-1},
\end{equation}
\begin{equation}
\label{N*}
N^*(T) \approx \left[{\pi\,\gamma(T)\,T\over 2\,d}\right]^{d\over d-1}\,T_{int}^{1\over d-1}\,\left[{2\over \nu\,(T-T_0)}\right]^{d+1\over d-1},
\end{equation}
and
\begin{eqnarray}
\label{alphaT}
\nonumber
\alpha(T)&\approx& (d-1)\left[{\pi\,\gamma(T)\over 2\,d}\right]^{d\over d-1}\left[{2\,(T\,T_{int})^{1\over 2}\over \nu\,(T-T_0)}\right]^{2\over d-1}\\\cr &\approx& {(d-1)\,\pi\,\gamma(T)\over 2\,d}\,R^*(T).
\end{eqnarray}
Thus we recover the Vogel-Fulcher formula for $\tau_{\alpha}(T)$ near $T=T_0$, with $\alpha(T) \propto (T-T_0)^{-1}$ for $d=3$.  

Note that, in Eq.(\ref{alphaT}), $\alpha(T)$ is linearly proportional to $R^*(T)$ independent of dimensionality $d$.  This formula and parts of the preceding analysis resemble recent results of Eckmann and Procaccia \cite{ECKMANN-PROCACCIA08}, who make a mathematically systematic study of a two-dimensional model and find both chainlike excitations and a linear relation between the super-Arrhenius activation energy and the spatial extent of the chains.   

Now return to Eq.(\ref{alphatilde}) and notice that, if $J(N) \approx c\,N^{d+1}$ for small $N$, this equation can be made to have the form
\begin{equation}
\tilde\alpha(T,N)\approx -\nu\,N\,\left(1-{T_A\over T}\right)
\end{equation}
by choosing 
\begin{equation}
c = {(2\,d)^d\over (d+1)^{d+1)}}\,{\bigl[\nu\,(T_A-T_0)\bigr]^{d+1}\over T_{int}\,\bigl[\pi\,\gamma(T_A)\,T_A\bigr]^d}.
\end{equation}
This choice of $J(N)$ means that $E_{int}$ cuts off rapidly for small chains, as it should, and that the contribution of the chains to the activation energy vanishes smoothly at $T_A$.  A simple way in which to interpolate between the small-$N$ and large-$N$ limits is to write
\begin{equation}
\label{J}
J(N)={c\,N^{d+1}\over 1 + c\,N^{d-1}}.
\end{equation} 
This interpolation formula will be used extensively in the following analyses of experimental data. 

It seems natural to assume that $\ell\,R^*(T)$ sets the length scale for the heterogeneous pattern of glassy domains and mobile regions discussed in Sec.\ref{CTRW}.  The XC picture implies that molecular configurations confined within domains smaller than $\ell\,R^*(T)$ are frozen because excitation chains are always subcritical at those length scales.  On the other hand, domains appreciably larger than $\ell\,R^*(T)$ can support XC-induced rearrangements and thus may be unstable against breaking up into smaller domains.  I therefore propose that the $R^*(T)$ emerging from the XC theory be the same as the the $R^*(T)$ in the CTRW analysis up to a dimensionless proportionality factor that, for present purposes, I set to unity. That is, I assume that this proportionality factor is absorbed into other quantities such as $\ell$ and the jump-length parameter $a$ introduced in Eq.(\ref{frjump}).

In \cite{Langer-PRE-06,Langer-PRL-06}, I argued that the diverging length scale $R^*(T)$ implies that the configurational entropy extrapolates toward zero at $T = T_0$, and does this in a way that is consistent with the apparent relation between the jump in the specific heat and the fragility near this point.  This thermodynamic part of the theory is interesting; but it is even more speculative than the dynamic part, and is not directly relevant to the main issues being addressed here. Nevertheless, it does provide a clue about how to evaluate the glassy and mobile fractions, ${\cal P}_G(T)$ and ${\cal P}_M(T)=1-{\cal P}_G(T)$. My guess in \cite{Langer-PRE-06,Langer-PRL-06} was that the excess configurational entropy might be proportional to the mobile fraction of molecules ${\cal P}_M(T)$, which in turn might be proportional to the surface-to volume ratio for glassy domains.  More specifically, I proposed that, for $d=3$, the excess entropy be proportional to $ 3\,h_B^*/R^*(T)$, where the length $h_B = \ell\,h_B^*$ is the thickness of the interdomain boundary.  To extrapolate this geometric relation to $T=T_A$, where $R^*$ and ${\cal P}_G$ vanish, I propose that 
\begin{equation}
\label{PGapprox}
{\cal P}_G(T) \cong \left({R^*(T)\over h_B^* + R^*(T)}\right)^d.
\end{equation} 
As will be seen, this simple extrapolation seems to work fairly well.

\begin{figure}[h]
\centering \includegraphics[height=7 cm]{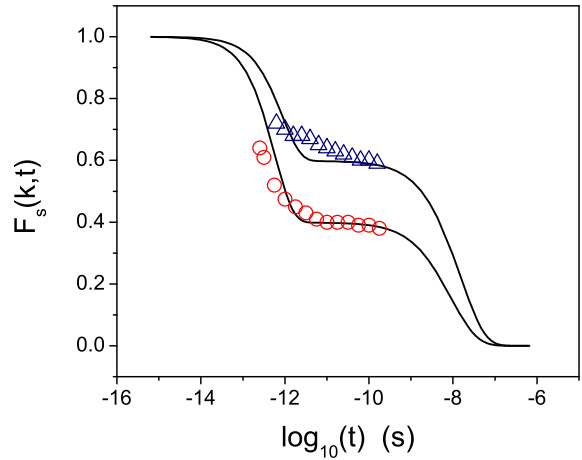}
\caption{\label{Fig-ISF-1}  (Color online) Intermediate scattering functions for $T=293\,K$, $k = 2\,\AA^{-1}$ (red circles) and $1.2\,\AA^{-1}$ (blue triangles).  The data points are taken directly from the graphs shown in Refs. \cite{Sillescu-etal-ISF-92,Sillescu-etal-ISF-93}. The solid curves are the theoretical results described in the text.}
\end{figure}

\section{Lessons from ortho-terphenyl}
\label{OTP} 

The most extensively studied laboratory glass is ortho-terphenyl (OTP), for which the viscosity \cite{LAUGHLIN+UHLMANN-72,CUKIERMAN-etal-73} and the diffusion constant \cite{EDIGER-etal-06} are known down to about the glass temperature $T_g \cong 240\,K$.  The self intermediate scattering function $\hat F_s(k,t)$ for this material is known from neutron scattering measurements down to $T=293\,K$ \cite{Sillescu-etal-ISF-92,Sillescu-etal-ISF-93,Tolle-01}, i.e. down to about its mode-coupling temperature $T_{MC}$ \cite{GOTZE91,GOTZE92}.  In the following paragraphs, I use these three sources of information as guides for estimating the model parameters introduced in the preceding Sections.  Although I do not refer explicitly to them, my analysis is informed by numerical simulations, especially those reported in \cite{KA95}.

\subsection{Scattering Data}

Start by looking at the scattering data, which provides preliminary information about the time and length scales.  A selection from the data of \cite{Sillescu-etal-ISF-92,Sillescu-etal-ISF-93} is shown in Fig.\ref{Fig-ISF-1} along with theoretical curves computed using Eqs.(\ref{ISFudef} - \ref{hatnj}).  As seen in the figure, the decoupling between the fast ($\beta$) and slow ($\alpha$) modes seems clear at the largest scattering wave vector, $k = 2\,\AA^{-1}$, but is not sharp at $k = 1.2\,\AA^{-1}$ or, as will be seen, at higher temperatures.  By necessity, I assume that $T \equiv T_s = 293\,K$ is far enough into the thermally activated, heterogeneous regime for the preceding CTRW analysis to be valid, at least at long enough times.

\begin{figure}[h]
\centering \includegraphics[height=7 cm]{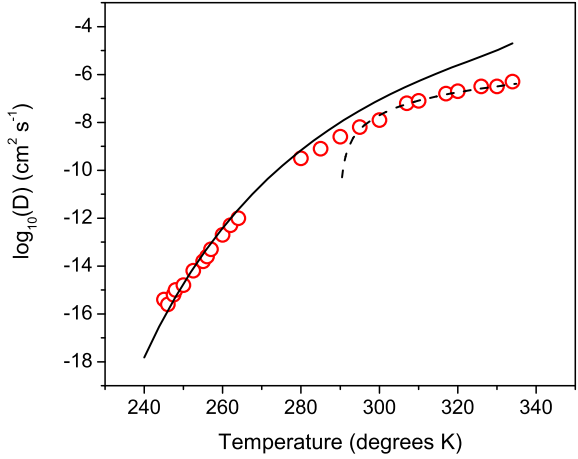}
\caption{\label{Fig-D} (Color online) Diffusion Constant $D_{\infty}(T)$.  The data points (red circles) are taken directly from the graphs shown in Ref. \cite{EDIGER-etal-06}. The solid curve is the theoretical fit obtained from the CTRW and XC theories.  The dashed curve is the mode-coupling approximation. }
\end{figure}

Eqs.(\ref{nMku}) and (\ref{Wdef}) imply that, for large $\Delta$, $\tilde n_M(k^*,u)$ effectively has a pole on the negative real axis at
\begin{equation}
\label{w0}
u \cong - \Delta\,[1- {\cal P}_M\,\hat f(k^*)]\equiv - w_0.
\end{equation}
At that point, $\tilde \psi_G(u)$ is negligibly small but has a nonzero imaginary part, so that the integrand obtained by closing the contour around the cut on the negative real $u$ axis has an approximate $\delta$-function at $u= -w_0$.  This piece of the inverse Laplace transform contributes a rapidly decaying -- i.e. ``$\beta$'' -- part of $\hat n_M(k^*,t^*)$:
\begin{equation}
\hat n_M^{(\beta)}(k^*,t^*)= A_M\,e^{- w_0\,t^*},~~A_M = {1-\hat f(k^*)\over 1- {\cal P}_M\,\hat f(k^*)}.
\end{equation}
The function $\tilde n_G(k^*,u)$ also has a pole at $-w_0$; but its residue is vanishingly small for large $\Delta$ because molecules frozen in glassy domains do not undergo prompt $\beta$ relaxation.  The plateau in $\hat F_s(k^*,t^*)$ starts at $t^*\cong t_{\beta}^* \cong 1/w_0$ and extends to $t^*\cong t_{\alpha}^* \sim 1$.  The height of the plateau is 
\begin{equation}
\label{Fplat}
\hat F_s^{plat} \cong {\cal P}_G +(1-A_M)\,{\cal P}_M \cong {{\cal P}_G\over 1-{\cal P}_M\,\hat f(k^*)}.
\end{equation}

The OTP scattering data for $T\equiv T_s =293\,K$ shown in Fig. \ref{Fig-ISF-1} suggest that $\hat F_s^{plat} \cong 0.4$ for $k = 2\,\AA^{-1}$ and (with much less certainty) $\hat F_s^{plat} \cong 0.6$ for $k = 1.2\,\AA^{-1}$.  Inserting these estimates into Eq.(\ref{Fplat}), I find ${\cal P}_M(T_s) \cong 0.75$ and $a\,R^*(T_s) \cong 0.60$.  As explained earlier, we cannot expect this CTRW analysis to describe $\beta$ relaxation accurately at small times, and, indeed, the data in Fig. \ref{Fig-ISF-1} do not exhibit a sharp fall-off at $t_{\beta}$. The best I can do is to estimate $t_{\beta}(T_s) \sim 10^{-12}$ s for $k = 2\,\AA^{-1}$.  Knowing ${\cal P}_M$ and $\hat f(k^*)$ at this point, I deduce from Eq.(\ref{w0}) that $\tau_M \sim 2 \times 10^{-14}$ s. Then, if $\tau_M$ is a thermal activation time of the form

\begin{figure}[h]
\centering \includegraphics[height=7 cm]{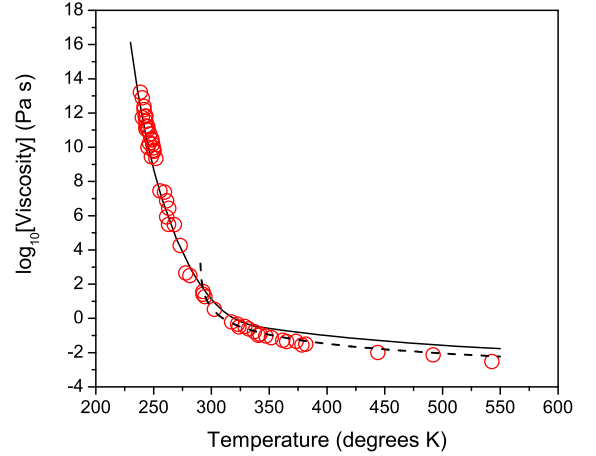}
\caption{\label{Fig-eta} (Color online) Newtonian viscosity $\eta_N(T)$. The data points (red circles) are those given in Refs. \cite{LAUGHLIN+UHLMANN-72,CUKIERMAN-etal-73}. The solid curve is the theoretical fit obtained from the STZ and XC theories.  The dashed curve is the mode-coupling approximation.}
\end{figure}
\begin{equation}
\label{tauMdef}
\tau_M= \tau_0\,\exp(T_M/T),
\end{equation}
and if $\tau_0 \cong 10^{-15}$ s, it follows that $T_M \sim 500\,K$.  Thus, $T= 293\,K$ is inside the region of thermally activated dynamics, but only marginally so.

The scattering data at $T_s=293\,K$, $k=2\,\AA^{-1}$, also imply very roughly that $t_{\alpha}^*/t_{\beta}^* \cong w_0\tau_{\alpha}^R\sim 10^3$.   Again, knowing ${\cal P}_M$ and $\hat f(k^*)$, I deduce that $t_{\alpha}/t_{\beta}\cong \tau_{\alpha}/\tau_M \cong \exp[T_Z/T_s+\alpha(T_s)]/\exp(T_M/T_s) \sim 10^3$, where $\alpha(T_s)$ is defined in Eq.(\ref{alphadef}). Thus, if $T_M \cong 500\,K$, then $T_Z \cong 2500\,K - T_s\,\alpha(T_s)$, which is substantially smaller than the value $T_Z \cong 3100\,K$ obtained in \cite{KTZ+96} by fitting the the viscosity data directly to an expression of the form of Eq.(\ref{alphadef}) -- a difference that turns out to be significant.  

\subsection{Diffusion and Viscosity Coefficients}

The next step is to find model parameters that fit the diffusion and viscosity data shown in Figs. \ref{Fig-D} and \ref{Fig-eta} respectively.  To evaluate the diffusion constant, start with the general formula for the mean-square displacement of a tagged molecule is
\begin{equation}
\langle {r^*}^2(t^*)\rangle = - 3\,\left[{\partial^2 \hat F_s(k^*,t^*)\over \partial {k^*}^2}\right]_{k^*=0},
\end{equation}
where $\hat F_s(k^*,t^*)$ is defined in Eq.(\ref{ISFtdef}). To evaluate this formula in the limit of large $t^*$, integrate around a small circle at the origin in the $u$ plane, and use the expansion $\tilde\psi_G(u)\approx 1 - 3u/2 + \cdots$.  The resulting integrand has a second-order pole at the origin, implying that
\begin{equation}
\langle {r^*}^2(t^*)\rangle \approx {3\,a^2\,t^*\over (3/2)\,{\cal P}_G + 1/\Delta}.
\end{equation}
Returning to dimensional units using Eqs.(\ref{tau*def}) and (\ref{Deltadef}),  and defining the asymptotic diffusion constant $D_{\infty}$ by $\langle r^2 \rangle \approx D_{\infty}\,t$, we find
\begin{equation}
\label{Dinfty}
D_{\infty}(T) = {3\,a^2\,\ell^2\over (3/2)\,{\cal P}_G(T)\,\tau_{\alpha}(T)+\tau_M(T)}.
\end{equation}
Thus, $D_{\infty} \propto \ell^2/\tau_{\alpha}$ at low temperatures where ${\cal P}_G \cong 1$ and $\tau_{\alpha} \gg \tau_M$.  When $T$ approaches or exceeds $T_A$, however, ${\cal P}_G$ vanishes and $D_{\infty} \propto \ell^2/\tau_M$.  

An analogous formula for the Newtonian viscosity $\eta_N(T)$ has been derived from the shear-transformation-zone (STZ) theory in \cite{JSL-STZ-PRE08}.  It has the form 
\begin{equation}
\label{etadef}
\eta_N(T) = \eta_0\,{T\over T_E}\,\left[\tau_{\alpha}'(T) + \tau_E(T)\right].
\end{equation}
This equation is the same as Eq.(7.10) in \cite{JSL-STZ-PRE08} except that the latter contains an incorrect and almost entirely irrelevant factor of 1/2.  In Eq.(\ref{etadef}), $\eta_0$ is a dimensional prefactor, $\tau_{\alpha}'(T)$ is the same as $\tau_{\alpha}(T)$ in Eq.(\ref{alphadef}) with $T_Z$ replaced by $T_Z'$; and
\begin{equation}
\tau_E(T) = \tau_0\,\exp\,\left({T_Z'+T_E\over T}\right).
\end{equation}
The quantities $k_B\,T_Z'$ and $k_B\,T_E$ are, respectively, the characteristic formation energy for STZ's and the Eyring activation energy for STZ transitions.

\subsection{Stokes-Einstein Violation}

The results of Mapes et al.\cite{EDIGER-etal-06} indicate a strong violation of the Stokes-Einstein relation; that is, the measured value of the product $D_{\infty}(T)\,\eta_N(T)/T$  is larger by a factor of about $10^2$ near $T_g$ than it is near and above $T_A$ (or $T_{MC}$), where it does remain constant as a function of $T$ as it should in a liquidlike material.  The conventional explanation for this behavior is that the low-temperature heterogeneities allow a diffusing molecule to move  further than it would in a homogeneous system. Such an enhancement of the diffusion length is built into the CTRW analysis in Sec.\ref{CTRW}, e.g. in Eq.(\ref{frjump}).  However, the effect is exactly cancelled out in Eq.(\ref{Dinfty}) by the correspondingly longer times that a molecule spends in glassy domains between diffusive jumps.  There is, of course, no reason to believe that the Stokes-Einstein  relation applies to the super-Arrhenius behavior of glass forming materials, where the mechanisms producing viscous deformation (localized shear transformations) and diffusion (delocalized hopping) are quite different from one another.

The preceding discussion of the characteristic temperature $T_Z$ suggests that the answer to the Stokes-Einstein question is that the $T_Z$ appearing in the diffusion analysis leading to Eq.(\ref{Dinfty}) is substantially smaller than the $T_Z'$ appearing in the viscosity formula, Eq.(\ref{etadef}). In other words, the formation energies for density fluctuations may be substantially smaller than those required for creating STZ's, which apparently are larger objects. (See the discussion in \cite{JSL-STZ-PRE08}, Sec.IV.) The excitation-chain analysis in Sec.\ref{XC} implies that the super-Arrhenius function $\alpha(T)$ and its ingredients are the same for both diffusion and viscosity; but there is no implication that the formation energies $T_Z$ and $T_Z'$ should be identical.  Note then that the estimate
\begin{equation}
{D_{\infty}(T_g)\,\eta_N(T_g)\over T_g} \approx {\tau_{\alpha}'(T_g)\over \tau_{\alpha}(T_g)} \approx \exp\left({T_Z'-T_Z\over T_g}\right) \sim 10^2
\end{equation}
implies that $T_Z'-T_Z \sim 1000\,K$, in rough agreement with the estimate made by comparing time scales for decoupling at $T \approx T_{MC}$.

\subsection{Parameters and Length scales}
 
These order-of-magnitude estimates provide a starting point for using the diffusion and viscosity data to determine the remaining model parameters.  Well within the super-Arrhenius region, the formulas for $D_{\infty}(T)$ and $\eta_N(T)$, in Eqs.(\ref{Dinfty}) and (\ref{etadef}) respectively, are most sensitive to the function $\alpha(T)$ which, in turn, is determined primarily by the choice of the Kauzmann/Vogel-Fulcher temperature $T_0$ and the product $T_{int}\,\gamma_0^3$. This sensitivity is best seen in the approximate formula for $\alpha(T)$, Eq.(\ref{alphaT}); $T_0$ moves the Vogel-Fulcher singularity and $T_{int}\,\gamma_0^3$ determines its strength.  I find that suitable values of $T_{int}\,\gamma_0^3$ are always in the neighborhood of $10^5\,K$.  In order that $T_{int}$ have roughly the same size as the formation energies $T_Z$ and $T_Z'$, we need $\gamma_0 \cong 3$.  I also find that $T_0 \cong 178\,K$.  

\begin{figure}[h]
\centering \includegraphics[height=7 cm]{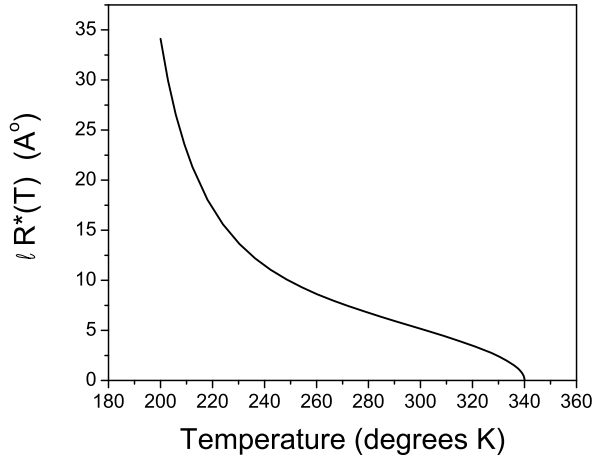}
\caption{\label{Fig-R} Length scale $\ell\,R^*(T)$, in Angstroms, as a function of temperature $T$, for the OTP parameters determined in the text.}  
\end{figure}

\begin{figure}[h]
\centering \includegraphics[height=7 cm]{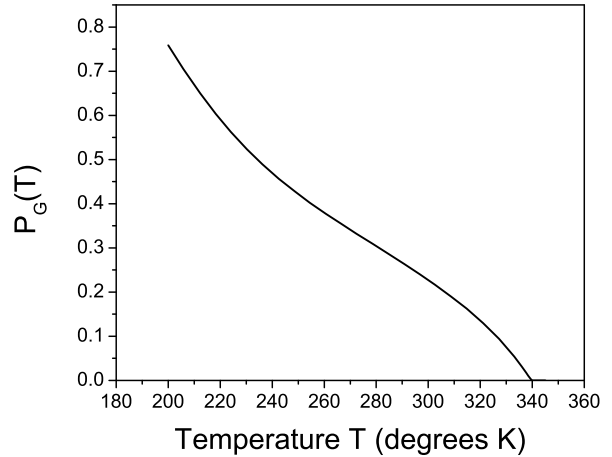}
\caption{\label{Fig-PG} Glassy probability ${\cal P}_G(T)$, as a function of temperature $T$, for the OTP parameters determined in the text.}  
\end{figure}

Accordingly, the theoretical curves in Figs. \ref{Fig-D} and \ref{Fig-eta} have been plotted using $T_0=178\,K$, $T_{int}\,\gamma_0^3 = 10^5$, and $\gamma_0 = 3$.  Other parameters are: $d = 3$, $\nu = \ln (6)$, $T_A=340\,K$, $T_M = 500\,K$, $T_E= 100\,K$, $\eta_0 = 6 \times 10^9$ Pa s, $T_Z = 2000\,K$ (for the diffusion constant), $T_Z'= 3000\,K$ (for the viscosity).  With these parameters, $\Delta(T) > 100$ for all $T < T_A$, consistent with the large-$\Delta$ assumption made in earlier estimates. From the discussion following Eq.(\ref{Fplat}), I find that the parameter determining the jump-length distribution in Eq.(\ref{frjump}) is $a = 0.074$.  The fit to the diffusion constant requires $a\,\ell = 0.05 \AA$; therefore $\ell = 0.7\,\AA$.   Finally, knowing that ${\cal P}_G(T_s = 293\,K) = 0.25$, I find from Eq.(\ref{PGapprox}) that $h_B = 4.7\,\ell \cong 3.3 \,\AA$.  Fig. \ref{Fig-R} shows a graph of $\ell\,R^*(T)$ in Angstroms as determined by these parameters; the corresponding  graph of ${\cal P}_G(T)$ is shown in Fig. \ref{Fig-PG}.

Note that $h_B$ is approximately half the diameter of an ortho-terphenyl molecule.  However,  $\ell$ is only about a tenth of this diameter, implying that the elementary links in excitation chains and the individual diffusive jumps in mobile regions are quite small, and therefore that those motions might involve cooperative rearrangements of many molecules rather than being simple, pairwise exchanges of positions. Interestingly, this value of $\ell$ is of the order of the Lindemann length, and perhaps is consistent with the idea that the excitation chains describe a local melting of the system during configurational rearrangements.  Other authors such as \cite{CHAUDHURIetal07,BERTHIER05}  have deduced similarly small length scales.  For example, using the parameters listed above and $T_g \cong 240\,K$, I find that $\ell\,R^*(T_g) \cong 16.4\,\ell \cong 11\,\AA$, or less than $2$ molecular diameters, roughly consistent with the results of \cite{BERTHIER05} (depending on -- among many other uncertainties -- whether $R^*$ is interpreted as the diameter or radius of a glassy domain). 

The theoretical fits to experimental data shown in these figures are interestingly imperfect.  In some cases, they can be improved cosmetically by adjusting parameters such as $T_M$, $T_Z$, etc., or by arbitrarily adjusting values of ${\cal P}_G$ for $T > 293\,K$ slightly downward from those determined by Eq.(\ref{PGapprox}).  Moreover,  if I relax the constraint that the product $T_{int}\,\gamma_0^3$ be the same for both diffusion and viscosity, and instead let this quantity be $0.8 \times 10^5$ for diffusion and $1.2 \times 10^5$ for viscosity, the theoretical curves for both $D_{\infty}(T)$ and $\eta_N(T)$ become indistinguishable from the data over their entire temperature ranges.  I doubt that such an adjustment is physically realistic.  To emphasize this point, in Figs. \ref{Fig-D} and \ref{Fig-eta}, I have added mode-coupling approximations of the form $D_{\infty} \propto (T-T_{MC})^2$ and $\eta_N \propto (T-T_{MC})^{-2}$, with $T_{MC}=290\,K$.  These simple approximations fit the high-temperature data extremely well, and illustrate the way in which the experimental results cross over at about $T_{MC}$ from solidlike, thermally activated behavior at low $T$ to liquidlike behavior at high $T$.

The discrepancies in fitting the scattering data are in part due to the same difficulty -- that the crossover region at temperatures near $T_{MC}$ seems theoretically intractible.  To illustrate this situation, I show in Fig. \ref{Fig-ISF-2} the scattering functions $\hat F_s(k,t)$ for $k = 2\,\AA^{-1}$ and temperatures $T = 327\,K,\,306\,K,293\,K$ (with accompanying experimental data) and $T= 280\,K$ (for which no data are available).  The two higher temperatures are clearly beyond the range of validity of the activation theory; the CTRW analysis is too blunt an instrument to resolve the details of the $\beta$ relaxation, and the activation barriers are too low.  Nevertheless, the agreement with experiment, even at $T=327\,K$, seems qualitatively reasonable, implying that the temperature and time scales may be at least approximately correct.  

\section{Stretched Exponential Relaxation}
\label{secSED}

\begin{figure}[h]
\centering \includegraphics[height=7 cm]{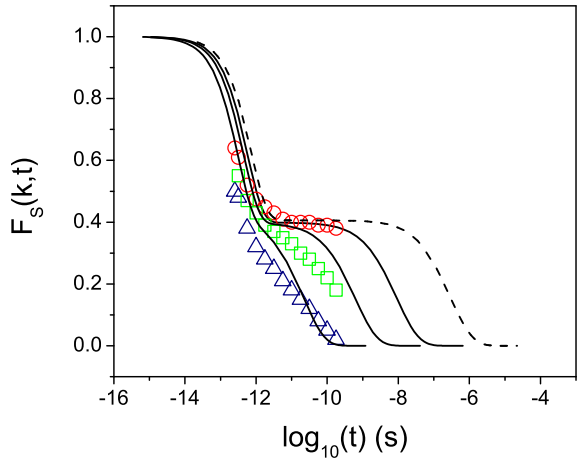}
\caption{\label{Fig-ISF-2} (Color online) Intermediate scattering functions for $k = 2\,\AA^{-1}$, $T=327\,K,\,306\,K,\,293\,K$, and $280\,K$ (dashed line), from left to right.}  
\end{figure}

Finally, it is useful to look more closely at the long-time tails of the scattering functions shown in Figs. \ref{Fig-ISF-1} and \ref{Fig-ISF-2}.  My interest here is not so much in the  experimental data for OTP as it is in the implications of the theory for interpreting extensions of that data. I find that the CTRW theory predicts stretched-exponential decay of $\hat F_s(k^*,t^*)$ with a temperature and wavenumber dependent index $b$ that crosses over to the intrinsic $b = 0.5$ only at unobservably long times $t^*$.  

\begin{figure}[h]
\centering \includegraphics[height=7 cm]{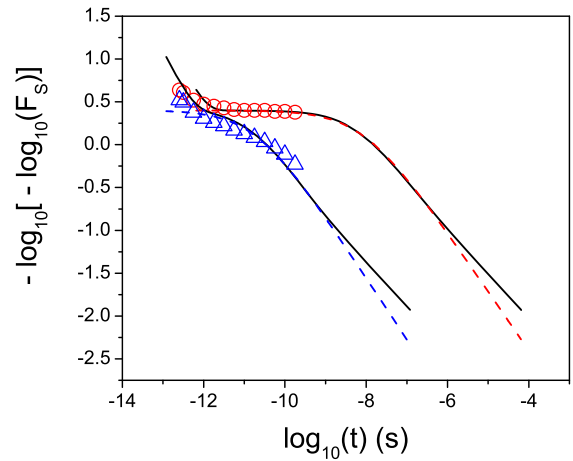}
\caption{\label{Fig-ISF-3} (Color online) Double logarithmic plot of  the intermediate scattering function for $k = 2\,\AA^{-1}$, for $T= 293\,K$ (upper black curve) and $T= 327\,K$ (lower black curve). The red and blue dashed curves are fits to pure stretched exponentials for the upper and lower curves with indices $b = 0.69$ and $0.73$ respectively.  The red circles and blue triangles are the same data points as those shown in  Fig. \ref{Fig-ISF-2}.}
\end{figure}

Especially in the case of the curve for $T= 293\,K$, $k = 2\,\AA^{-1}$, the CTRW analysis should be accurate at large $t^*$.  I have replotted that curve in Fig. \ref{Fig-ISF-3}, here in the form $-\log_{10}[- \log_{10}\hat F_s(k,t)]$ as a function of $\log_{10}(t)$, so that the slope is asymptotically equal to (minus) the stretched-exponential index $b$.  For comparison, I also have plotted the analogous curve for $T=327\,K$, again with $k = 2\,\AA^{-1}$, although that temperature is well into the mode-coupling region according to Figs. \ref{Fig-D} and \ref{Fig-eta}.  The experimental points are shown for both cases.

These curves, and others like them (not shown here) for different values of $k$ and $T$, exhibit many of the features seen in Fig. 2 of Ref \cite{JSL-SM-08}.  (The dimensionless wavenumber $k^*$ was denoted simply by $k$ in that paper.)  The analysis in \cite{JSL-SM-08} tells us that, in the limit of small $k^*$ at fixed large $t^*$, the behavior becomes diffusive, $b \to 1$.  Conversely, at fixed nonzero $k^*$, in the limit of large $t^*$, $b \to 0.5$.  Fig. 2 of \cite{JSL-SM-08} also shows that the apparent value of $b$ changes continuously from $b = 0.5$ at large $k^*$ to $b=1$ at small $k^*$.  

These observations provide a clue about what to expect in the temperature dependent theory.  Because $k^* = k\,\ell\,R^*(T)$, the wavenumber $k^*$ in \cite{JSL-SM-08} is a proxy for the temperature.  At fixed scattering wavenumber $k$, $k^*$ decreases as $T$ increases.  If $b$ approaches unity as $k^*$ decreases toward zero in the $T$-independent theory, then we may expect that, at fixed $k$, $b$ will behave in the same way with increasing $T$ as $R^*(T)$ becomes small.  Apparently this is what happens; but the temperature dependence of $b$ turns out to be very slow at low $T$, and the interesting high-temperature behavior is beyond the limits of validity of the present theory.

I have pursued this analysis in Fig. \ref{Fig-ISF-3} by comparing the CTRW results with pure stretched-exponential functions of the form
\begin{equation}
\label{SED}
\hat F_s^{SE}(k^*,t^*) = \hat F_s^{plat}\exp\,(- C_{SE}\,{t^*}^b), 
\end{equation}
where $C_{SE}$ and $b$ are fitting parameters and $\hat F_s^{plat}$ is determined by Eq.(\ref{Fplat}).  For the upper curve, $T=293\,K$, $C_{SE}= 0.75$, and $b=0.70$.  For the lower curve, $T=327\,K$, $C_{SE}=0.65$ and $b = 0.72$.  The fits seem to be exact all the way from the beginning of the plateau out to the point where the slope changes to $b = 0.5$ several decades in time later.  At this crossover, in both cases, $\hat F_s$ has dropped from about $0.4$ to an unobservably small value of the order of  $10^{-5}$.  To see the intrinsic $b = 0.5$, or the value of $b$ appearing in a generalized form of the waiting-time distribution $\psi_G(t^*)$ discussed following Eq.(\ref{psiG}), we would have to go well beyond this crossover point.  Thus, any attempt to deduce $b$ from an experimental curve like this would produce an intermediate value, equal in this case neither to unity nor to $0.5$.  

Further comparisons indicate that the effective index $b$ deduced in this way increases slowly with increasing $T$. The value of $b$ at $T=327\,K$ is slightly larger than that at $T=293\,K$.  At the higher $T$, $b$ is measured over a smaller range of times and therefore with less precision; but the small difference between the $b$ values is within the accuracy of the fits.  However, at a lower temperature, $T= 280\,K$, the parameters $b$ and $C_{SE}$ are the same as those at $293\,K$ to the accuracy of my calculation.  $b$ does depend more strongly on $k$ at fixed $T$.  At  $k = 1.2\,\AA^{-1}$ and $T=293\,K$, I find $b = 0.77$ -- appreciably larger than at $k = 2\,\AA^{-1}$.  I also have checked to make sure that the value of $b$ measured by this fitting procedure is always unambiguously equal to unity if $\psi_G(t^*)$ is a simple exponential function. 

To see what is happening here in somewhat more detail, look at an approximation for $\hat F_s(k^*,t^*)$ that is accurate for $\Delta(T)\gg 1$ and for values of $t^*$ large enough that we can use the  small-$w$ approximation, $\tilde\psi_G(-w \pm i\,\epsilon)\approx 1 + 3 w/2 + \cdots \mp i\,B(w)$, where $B(w)$ is defined in Eq.(\ref{A+iB}).  I find
\begin{eqnarray}
\label{Fasymp}
\nonumber
&&\hat F_s(k^*,t^*)\cong {w_A(k^*,T)\over \hat f(k^*)}\int_0^{\infty} {dw\over \pi\,w}\\ \cr
&&\times {(2/3)\,B(w)\,\exp(-w\,t^*)\over \bigl[w - w_A(k^*,T)\bigr]^2 + \bigl[(2/3)\,B(w)\bigr]^2},
\end{eqnarray}
where 
\begin{equation}
\label{wAdef}
w_A(k^*,T)= {2\,\bigl[1-\hat f(k^*)\bigr]\over 3\,{\cal P}_G(T)\,\hat f(k^*)}.
\end{equation}
Apart from the prefactor $\hat f(k^*)^{-1}$, which does not affect $b$, all of the temperature and wavenumber dependence is contained in $w_A(k^*,T)$.  That is, the effective $b(k,T)$ is a function of only $w_A$ within the range of validity of Eq.(\ref{Fasymp}).  

In the limit of vanishingly small $k^*$, and at large fixed $t^*$, $\hat f(k^*)\approx 1$, $B(w_A) \ll 1$, and the integrand in Eq.(\ref{Fasymp}) is accurately approximated by a $\delta$ function at $w = w_A \propto {k^*}^2$.  Therefore
\begin{equation}
\hat F_s(k^*,t^*) \approx e^{-w_A\,t^*} = \exp\left[-{2\,a^2\,\ell^2\,k^2\,t\over 3\,{\cal P}_G(T)\,\tau_{\alpha}(T)}\right],
\end{equation}
which confirms that $b \to 1$ in the small-$k^*$ limit. 

At $k=2\,\AA^{-1}$ and $T=293\,K$, however, $w_A \cong 2.8$; therefore, for times $t^*$ of the order of unity or greater, the integrand in Eq.(\ref{Fasymp}) is negligibly small at $w \sim w_A$.  At values of $t^*$ so large that the integrand is dominated by the product $B(w)\,\exp(-w\,t^*) \propto \exp(-w\,t^*-1/w)$, there is a sharp peak at $w = 1/\sqrt{t^*}$, and a saddle-point approximation yields
\begin{equation}
\label{saddle}
\hat F_s(k^*,t^*)\propto 2\,\sqrt{t^*}\,e^{-2\,\sqrt{t^*}}.
\end{equation}
At large $t^*$, this function is essentially the same as $\phi_G(t^*)$ given in Eq.(\ref{phiG}).  However, the approximations made in obtaining this result are not valid until the saddle point is well below the maximum of $B(w)$ at $w = 2/3$.  Therefore, the saddle must occur at values of $w$ less than about $0.3$, and Eq.(\ref{saddle}) is correct only for $t^* \ge 10$, i.e. $t \ge 10^{-7}$ s, which is roughly where the crossover from $b = 0.70$ to $b = 0.5$ occurs on the upper curve in Fig. \ref{Fig-ISF-3}.    

The remarkable feature of these numerical results is that the scattering function $\hat F_s(k^*,t^*)$ predicted by the CTRW theory is so well fit by a simple stretched-exponential in its experimentally observable domain. I have checked that Eq.(\ref{Fasymp}) accurately reproduces the large-$t^*$ behavior of $\hat F_s(k^*,t^*)$ in both the effective $b(k,T)$ regime and the asymptotic $b = 0.5$ limit, but can find no hint from this formula that Eq.(\ref{SED}) is anything more than a simple -- and therefore non-unique -- two-parameter fit to a smooth function that decreases more slowly than exponentially in a restricted range of values of $t^*$.  The analysis does make it clear that stretched-exponential behavior depends on there being an intrinsic stretched-exponential waiting time distribution $\psi_G(t^*)$.  The important outstanding question is whether the intrinsic $b$ itself must be temperature dependent in order that the observed index $b$ exhibit physically realistic temperature and wavenumber dependence. 

\section{Concluding Remarks}
\label{conclusions}

The introduction to this paper begins with a list of questions about the role of spatial heterogeneities in the dynamics of glass-forming liquids.  Several of those questions were addressed and at least partially answered in the preceding paper \cite{JSL-SM-08}; others are discussed in this one.  I conclude here by summarizing my present opinions about each of them.

{\it Structural or dynamic heterogeneity?}  I think that this is largely a semantic issue.  We referred in \cite{JSL-SM-08} to papers  by Widmer-Cooper and Harrowell \cite{HARROWELL}, who showed that molecules in mobile regions -- regions of high ``propensity'' -- have smaller than average Debye-Waller factors, meaning that they are participants in elastically soft modes.  Such modes are ultimately structural in nature, but are to some degree delocalized and not easily detected by looking at near-neighbor structural correlations.  I think  that the mobile regions are the locations of these soft modes, and that the nonlinear $\beta$ rearrangements occur predominantly in these places where the linear elastic modes are softest and most highly excited by thermal fluctuations.  The fact that the pattern of mobile regions and glassy domains is fluctuating on $\alpha$ time scales adds an extra dynamic element to the picture of the heterogeneous system as a whole.  

{\it Heterogeneity and non-Gaussian displacement distributions?} The conclusion reached in \cite{JSL-SM-08}, consistent with the results of \cite{CHAUDHURIetal07}, is that non-Gaussian displacement distributions are robust, mathematical features of essentially all heterogeneous diffusion models in which there is strong decoupling between the fast and slow modes.  This result depends only on the requirement that $\psi_G(t^*)$ decays rapidly enough that $\tilde \psi_G(u)$ is very small at large $u$; it does not depend at all on whether $\psi_G(t^*)$ is a stretched exponential function.  I have not repeated this analysis here because, apart from some inconsequential technical differences, it is exactly the same as in Sec. V of \cite{JSL-SM-08}.  Note that, while stretched exponentials and non-Gaussian distributions are both generated by spatial heterogeneities, these are independent phenomena; non-Gaussian diffusion can occur in the absence of stretched exponential relaxation.  

{\it Violation of Stokes-Einstein?} Although it might have been satisfying to find a theoretically more elegant explanation -- longer diffusion paths or a breakdown of ergodicity -- I argue in Sec. \ref{OTP}-C  that the violation of the Stokes-Einstein relation has a more prosaic origin.  Specifically, as a glass-forming liquid becomes more solidlike below $T_A$, the localized shear transformations producing viscous deformation become less and less like the delocalized hopping mechanisms responsible for diffusion.  Looked at from this point of view, the large violation of Stokes-Einstein is a prediction of the shear-transformation-zone (STZ) theory \cite{JSL-STZ-PRE08}, and the scale of the low-$T$ violation (the factor of 100 found in \cite{EDIGER-etal-06}) is independently predicted by the high-$T$ time-scale analysis described in Sec. \ref{OTP}-A.
 
{\it Length scales?} I cannot prove that the estimates of relatively small length scales in Sec. \ref{OTP}-D are unique interpretations of the available data, but I suspect that the conclusions are at least qualitatively correct.  If so, it seems especially urgent to understand what is happening here.  What kinds of molecular rearrangements correspond to such small net displacements?  On the other hand, there are still substantial  uncertainties.  For example, the poorly understood parameter $\gamma_0$ plays an important role in these estimates, and the quantity $R^*(T)$ that appears as the critical size of an excitation chain in Sec. \ref{XC} may not be exactly the same as the size of the glassy domains in Sec. \ref{CTRW}. The most serious experimental uncertainties pertain to the scattering data.  It would be useful if these could be extended to lower temperatures and longer times.

{\it Heterogeneity and stretched exponentials?} The connection between heterogeneity and stretched exponentials presented in \cite{JSL-SM-08} seems plausible; but it is only one step more specific than the conventional picture in which a stretched-exponential function is just an incoherent superposition of ordinary exponential decays occurring in different parts of a heterogeneous system. I think it may be possible to make a first-principles estimate of the intrinsic, possibly $T$-dependent, stretched-exponential index $b$ that appears in the glassy, waiting-time distribution $\psi_G(t^*)$.  However, the discussion in Sec. \ref{secSED} implies that this intrinsic index is impossible to measure directly, and that further calculations are needed in order to make contact with experiments.  The positive aspect of the situation is that the theoretical correlation functions, like many seen in experiments and numerical simulations, exhibit sharply defined, temperature and wavenumber dependent indices $b(k,T)$ over long times.  Those indices should be calculable from the intrinsic index $b$ via CTRW analyses such as the one described here.

\begin{acknowledgments}
This research was supported by U.S. Department of Energy Grant No. DE-FG03-99ER45762. 
\end{acknowledgments}

\end{document}